\documentclass[9.5pt,lettersize,journal]{IEEEtran}
\usepackage{amsmath,amssymb,amsthm}
\usepackage{algorithm}
\usepackage{algorithmic}
\usepackage{array}
\usepackage[caption=false,font=normalsize,labelfont=sf,textfont=sf]{subfig}
\usepackage{textcomp}
\usepackage{stfloats}
\usepackage{url}
\usepackage{verbatim}
\usepackage{graphicx}
\usepackage{booktabs}
\def\BibTeX{{\rm B\kern-.05em{\sc i\kern-.025em b}\kern-.08em
    T\kern-.1667em\lower.7ex\hbox{E}\kern-.125emX}}
\usepackage{balance}

\newtheorem{thm}{Theorem}[section]
\newtheorem{lem}[thm]{Lemma}

\newtheorem{cor}{Corollary}

\theoremstyle{definition}
\newtheorem{defn}{Definition}[section]

\begin{document}
\title{Differentially Private Distance Query with Asymmetric Noise}
\author{
        Weihong~Sheng,
        Jiajun~Chen,
        Chunqiang~Hu,
        Bin~Cai,
        Meng Han,
        Jiguo Yu
\IEEEcompsocitemizethanks{\IEEEcompsocthanksitem Weihong Sheng, Jiajun Chen, Chunqiang Hu and Bin Cai are with the School of Big Data \& Software Engineering, Chongqing University, Chongqing 400044, China. E-mail: \{swh, jiajunchen, chu, caibin\}@cqu.edu.cn.
\IEEEcompsocthanksitem Meng Han is with Binjiang Institute, Zhejiang University. Email: mhan@zju.edu.cn.
\IEEEcompsocthanksitem Jiguo Yu is with Big Data Institute, Qilu University of Technology, Jinan, Shandong 250353, China (e-mail: jiguoyu@sina.com).
}

}

\maketitle

\begin{abstract}
With the growth of online social services, social information graphs are becoming increasingly complex. Privacy issues related to analyzing or publishing on social graphs are also becoming increasingly serious. Since the shortest paths play an important role in graphs, privately publishing the shortest paths or distances has attracted the attention of researchers. Differential privacy (DP) is an excellent standard for preserving privacy. However, existing works to answer the distance query with the guarantee of DP were almost based on the weight private graph assumption, not on the paths themselves. In this paper, we consider edges as privacy and propose distance publishing mechanisms based on edge DP. To address the issue of utility damage caused by large global sensitivities, we revisit studies related to asymmetric neighborhoods in DP with the observation that the distance query is monotonic in asymmetric neighborhoods. We formally give the definition of asymmetric neighborhoods and propose Individual Asymmetric Differential Privacy with higher privacy guarantees in combination with smooth sensitivity. Then, we introduce two methods to efficiently compute the smooth sensitivity of distance queries in asymmetric neighborhoods. Finally, we validate our scheme using both real-world and synthetic datasets, which can reduce the error to $0.0862$.

\end{abstract}

\begin{IEEEkeywords}
Differential privacy, asymmetric neighborhood, shortest path, distance, graph.
\end{IEEEkeywords}

\section{Introduction} \label{sec:1}
\IEEEPARstart{W}{ith} the rise of privacy concerns, more and more data analysis or publishing tasks require enhanced privacy preservation to avoid ethical or legal issues. Traditional data anonymization techniques, such as $k$-anonymity \cite{liu2008towards}, $l$-diversity \cite{aggarwal2008general}, etc., are unable to effectively deal with increasingly powerful attackers \cite{narayanan2009anonymizing} because these techniques rely on accurate segmentation of sensitive attributes and quasi-identifiers. Differential privacy (DP) \cite{dwork2006differential}, which has gained the favor of researchers in the field of privacy, has become a gold standard of privacy preservation. DP addresses the limitations of traditional data anonymization techniques, providing robust defense against attacks such as linking attacks and differential attacks. DP remains resilient against powerful attackers, even when these attackers possess all knowledge except the target record. Its applicability extends to various data analysis and publishing tasks, with a notable emphasis on complex graph analysis, which constitutes a significant research area.

The type of graph data pervades all aspects of social life, such as traffic graphs, citation graphs, and social graphs. With the improvement of basic network facilities and the popularization of mobile smart terminals, graph data have become increasingly complex and diversified. Thus, increasingly complex social information graphs have been the hardest hit by privacy breaches due to their close relevance to humans \cite{li2017privacy,cai2020privacy,wang2023anchor}. In social information graphs, sensitive information can be weights, the existence of edges or vertices, and statistical information (degree distribution, triangle counts, and clustering coefficients, etc.). With different analysis or publishing tasks, DP has been widely used in graphs \cite{imola2021locally,jian2021publishing,291046}.

The shortest path is an important metric for graphs and is often used in route scheduling. The shortest path query, which obtains the shortest path from source to target, is the basic operation for route scheduling. To avoid confusion, we emphasize that in this paper, 'shortest path' refers to the shortest path itself, which is an edge sequence, while 'distance' refers to the length of the shortest path. Friend matching is also a route scheduling task to find the shortest path from source user to target user in social graphs. With privacy concerns, how to privately answer the shortest path or distance query is a critical issue in the social graphs.

Shortest path and distance publishing under DP was first formally studied by Sealfon \cite{sealfon2016shortest}. This work is built on the weight private graph assumption. That is, the topology is public, but the edge weights are private information. Specifically, in the assumption, two graphs are considered to be adjacent if the total of their weights differs by no more than one. The following works have taken this assumption similarly to optimize the error \cite{fan2022distances,fan2022private,chen2023differentially}.

However, the weight private graph assumption is not a universal assumption. Private information in graphs can also be vertices or edges. Specifically, the edge represents a relationship between two users in a social graph. We can ignore the exact value of the edge weights, as long as that edge exists, it represents that there must be some relationship between the users. The relationship is the privacy that users do not want to expose (edge DP). To publish distances with DP, existing works provide two routes: adding noise to edge weights \cite{fan2022private} or adding noise to distance \cite{sealfon2016shortest}. But for unweighted graphs, adding noise to the distance is a better choice. Intuitively, since we reject the weight private graph assumption, the global sensitivity of the distance query is not $O(1)$ but $O(n)$ where $n$ is the number of vertices in the graph. This sensitivity will produce a significant amount of noise, leading to a catastrophic reduction in utility.

The large global sensitivity in complex social graphs is the main culprit that prevents DP from being applied to graphs. Relevant efforts are made to adopt means to reduce sensitivity \cite{imola2021locally,raskhodnikova2016lipschitz}. But these techniques do not work for distance publishing. Driven by the asymmetric neighbors used in One-sided Differential Privacy (OSDP) \cite{kotsogiannis2020one} and Asymmetric Differential Privacy (ADP) \cite{takagi2022asymmetric}, we can observe that the asymmetric neighbors can contribute to improving the utility of some queries by using exponential noise instead. Fortunately, the distance query satisfies the requirement of monotonicity. We give an example to describe the monotonicity.

Given a graph $G=(V,E)$, $G^\prime$ is its neighbor by adding an edge to $G$. Given the distance query $f_G(u,v)$ that returns the distance between the vertices $u$ and $v$, we have $f_G(u,v) \geq f_G^{\prime}(u,v)$ for any $u,v \in V$. If neighbors are generated by removing an edge, a similar result holds: $f_G(u,v) \leq f_G^{\prime}(u,v)$ for any $u,v \in V$.

With the above observation, we can improve the utility of the distance query by constraining the neighborhood to be asymmetric. The asymmetric means that, if $G^\prime$ is the neighbor of $G$, $G$ is not the neighbor of $G^\prime$. They are not exchangeable, which is different from the symmetry in \cite{chen2021edge}. We will formally analyze the 'asymmetric' in Sections \ref{sec:4} and \ref{sec:5}.

In this paper, to answer the distance query privately with improved utility, we revisit the asymmetric neighborhood setting in DP. We formally define the asymmetric neighbors and monotonicity. We emphasize that monotonicity is derived from the query function in a particular neighborhood, rather than sensitivity. We improve the ADP mechanism in \cite{takagi2022asymmetric} by introducing smooth sensitivity. To further improve the utility, we analyze the local sensitivity and the smooth sensitivity of the distance query in two different neighborhoods: adding an edge and removing an edge. We also propose two algorithms to calculate the smooth sensitivity and privately answer the distance query.

The main contributions are listed below.
\begin{itemize}
    \item Motivated by the study of previous works on asymmetric neighborhoods in DP, we formally generalize asymmetric neighborhoods and propose the Individual Asymmetric Differential Privacy (IADP).
    \item For two different neighborhoods, we analyze the complexity of global sensitivity and local sensitivity in distance queries. Then we further propose the expressions and efficient computation methods for the smooth sensitivities in the two neighborhoods.
    \item We present two algorithms to answer distance queries and illustrate their effectiveness with both real-world and synthetic datasets.
\end{itemize}

The remainder of this paper is organized as follows. Section \ref{sec:2} provides a review of related work. Section \ref{sec:3} presents the theoretical basis relevant to the proposed algorithms. Section \ref{sec:4} reviews the asymmetric property used in previous works. Section \ref{sec:5} gives our definition of the asymmetric neighborhood and the mechanism. Section \ref{sec:6} presents our solutions to publish distance privately. Section \ref{sec:7} demonstrates the results of our two solutions. Section \ref{sec:8} discusses the potential issues of our work. Section \ref{sec:9} gives a summary of this work and outlines our future work.

\section{Related Work}\label{sec:2}
With the rise in popularity of graphs in data analytics, researchers have begun to focus on privacy issues in graphs. Privacy in graphs can be basically categorized into edge privacy \cite{hay2009accurate}, vertex privacy \cite{raskhodnikova2016differentially}, and weight privacy \cite{sealfon2016shortest}. In edge privacy, edges are private information that should be masked. Imola et al. \cite{imola2021locally} consider the edge as privacy in a local setting and use graph projection to reduce the sensitivity of the triangle count query. Chen et al. \cite{chen2021edge} proposed an edge DP based mechanism to release the algebraic connectivity of graphs without breaching the privacy of connection. In this analysis, the sensitivity is a constant independent of the number of vertices $n$ of the graph. Similarly, vertex privacy treats the vertex as private and attempts to hide the existence of the vertex itself. Jian et al. \cite{jian2021publishing} proposed two node DP based algorithms to answer all graph queries. However, they only examine their solutions by the number of nodes and edges, the amount of triangles, and the clustering coefficient, which cannot cover all graph queries and provide the necessary utilities. For weight privacy, Zhou et al. \cite{zhou2019edge} provided a weight privacy-based spectrum query algorithm and proved its global sensitivity.

Currently, multiple works on shortest path or distance publishing with differential privacy follow the weight private graph assumption. Sealfon \cite{sealfon2016shortest} first used this assumption to privately release all-pair distances without concern for global sensitivity, since global sensitivity is a unit. In this work, the additional error of the approximate distance is at most $O(n \log n / \varepsilon)$. To further minimize this error, Fan et al. \cite{fan2022distances} proposed two approaches for distance publishing in tree and grid graphs, achieving errors of $O\left(\log ^{1.5} n\right)$ and $\tilde{O}\left(n^{3 / 4}\right)$, respectively. With the same assumption, Fan et al. \cite{fan2022private} realized privacy-preserving distance queries by constructing synthetic graphs with an error of $\tilde{O}\left(n^{1 / 2}\right)$. Chen et al. \cite{chen2023differentially} followed the same assumption, given a distance release algorithm with additional error $\tilde{O}\left(n^{2 / 3} / \varepsilon\right)$. The authors then emphasize that the lower error bound is at least $\Omega\left(n^{1 / 6}\right)$. Moreover, for bounded weights, this work improved the error to approximately $n^{(\sqrt{17}-3) / 2+o(1)} / \varepsilon$. Deng et al. \cite{deng2023differentially} followed the similar assumption, proposed a differentially private range query on shortest paths that achieve additive error with $\widetilde{O}\left(n^{1 / 3}\right)$ for $\varepsilon$-DP and $\widetilde{O}\left(n^{1 / 4}\right)$ for $(\varepsilon, \delta)$-DP. Cai et al. \cite{10308714} modified the assumption to a large weight difference. In this setting, the weight range is $[a,b]$ and the neighbors differ at most in $b-a$. Thus, they focused on the shortest path change rate, since the large sensitivity will destroy the utility of shortest path and distance queries.

The previous works have basically been fundamentally concerned with the shortest path and distance release problems on the weighted graphs, since they do not have to concern themselves with utility issues arising from a large sensitivity. However, there is a gap on how to privately publish the shortest paths or distances on unweighted graphs, where the large sensitivity prevents the direct application of the previous results.

\section{Preliminaries}\label{sec:3}
\subsection{Differential Privacy in Graph}

Let us consider a simple, connected and unweighted graph $G=(V,E)$, with $V$ the set of vertices and $E$ the collection of edges in $G$, respectively. We use $n,m\in \mathbb{N}$ to denote the number of vertices and edges. For simplicity, we use $[n]$ as the discrete set of $\{1,2,...,n\}$. With $V=\{v_1,v_2,...,v_n\}$, an edge $(v_i,v_j)$ exists if $v_i$ and $v_j$ are neighbors of each other for $i,j\in [n]$. Let $\mathcal{G}$ be the domain of all possible graphs, and let $f$ be the distance query function.

With the difference in the definition of the neighborhood, the DP in the graph has two types: the edge DP and the node DP\cite{hay2009accurate}. In edge DP, neighbors differ by one edge. Similarly, in the node setting, neighboring objects differ by one node and its associated edges. We neglect the node setting due to the similarity and introduce the edge DP below.

\begin{defn}
(Edge Neighboring) Given graphs $G = (V,E)$ and $G^\prime = (V,E^\prime)$, we say $G$ and $G^\prime$ are edge neighboring if they differ in at most a single edge:
\begin{equation}
    |E\bigtriangleup E^\prime| \leq 1 .
\end{equation}
where $\bigtriangleup$ refers to symmetry difference.
Thus, we denoted $G$ and $G^\prime$ as $G \sim G^\prime$.
\end{defn}

\begin{defn}
(Edge Differential Privacy) Let $M: \mathcal{G} \rightarrow \mathcal{O}$ be a randomized algorithm. For any input graphs $G$,$G^\prime \in \mathcal{G}$ if $G \sim G^\prime$, and for all possible outputs $O$, have:
\begin{equation}
    \operatorname{Pr}[\mathcal{M}(G) \in O] \leq \exp (\varepsilon) \times \operatorname{Pr}\left[\mathcal{M}\left(G^{\prime}\right) \in O\right] + \delta.
\end{equation}
We say that $M$ satisfies $(\varepsilon, \delta)$-DP or $(\varepsilon, \delta)$-Approximate DP where $\varepsilon$ and $\delta$ measure the level of privacy preservation of $M$. Specifically, $\varepsilon$ constrains the similarity of the distribution of $\mathcal{M}(G)$ and $\mathcal{M}(G^\prime)$, while $\delta$ represents the probability that the constraint fails. When $\delta = 0$, we say that $M$ satisfies $\varepsilon$-DP, which is also called pure differential privacy.
\end{defn}
With the guarantee of differential privacy, adversaries are prevented from learning enough information to distinguish which dataset results in the output $O$. This guarantee is reinforced by introducing randomness into the output, which constrains the difference originally caused by the neighboring edge within $\varepsilon$ during the process of $\mathcal{M}$. For example, consider edge $(v_i,v_j)$ as a relationship that needs to be urgently hidden in a social network. With DP, adversaries cannot increase their confidence in determining whether $v_i$ and $v_j$ are neighbors based on knowledge obtained from $\mathcal{M}(G)$ and $\mathcal{M}(G^\prime)$.

\subsection{Sensitivity}
A basic and general implementation of DP is adding Laplace noise to the query result if the query function is real-valued.
\begin{defn}
(Laplace Mechanism) Let the query function $f:\mathcal{G} \rightarrow \mathbb{R}^+$. $\mathcal{M}(G)$ satisfies $\varepsilon$-DP if we add Laplace noise which calibrated to the sensitivity of $f$:
\begin{equation}
    \mathcal{M}(G):=f(G)+\frac{GS_f}{\varepsilon} \cdot X ,
\end{equation}
where $X$ is a Laplace noise drawn from the Laplace distribution with probability density function $Lap(x)=\frac{1}{2}e^{-|x|} $.
\end{defn}

\begin{defn}
(Global Sensitivity) For any query function $f:\mathcal{G}\rightarrow \mathcal{R}$ where $\mathcal{G}$ is the domain of input graphs, with $G \sim G^\prime$, we have:
\begin{equation}
    GS_f :=\sup _{G \sim G^\prime}\left\|f(G)-f\left(G^{\prime}\right)\right\|_1 .
\end{equation}
\end{defn}
The global sensitivity is the property of the query function $f$ which is not related to a specific graph $G$. More specifically, it refers to the upper bound of $f(\mathcal{G})$. With this sensitivity, we can constrain what adversaries learn by calibrating the magnitude of noise.

Some query functions have low global sensitivity, e.g., degree distribution query, etc. However, some functions have high global sensitivity, e.g., triangle counting in graph, etc. The high global sensitivity will yield high noise, which can completely destroy the utility of the query results. Intuitively, the query service provider just needs to focus on the own dataset and its neighbors, without considering the universe.
\begin{defn}
(Local Sensitivity) Let $f:\mathcal{G} \rightarrow \mathbb{R}^+$. Given a dataset $G$, the local sensitivity of $f$ at $G$ is:
\begin{equation}
    LS_f :=\max _{G \sim G^\prime}\left\|f(G)-f\left(G^{\prime}\right)\right\|_1 .
\end{equation}
\end{defn}
Although $LS_f$ can reduce the magnitude of noise with the observation $GS_f \leq LS_f$, the privacy leakage may occur since the noise scale is closely related to the dataset.

\begin{defn}
(Smooth Sensitivity) Let $f:\mathcal{G} \rightarrow \mathbb{R}^+$. For $\beta > 0$ and $d(\cdot)$ a distance measure (Hanming Distance), we have smooth sensitivity of $f$ at $G$:
\begin{equation}
    SS_{f}(G):=\max _{G^\prime \in \mathcal{G}}\left(L S_f(G^\prime) \cdot e^{-\beta d(G, G^\prime)}\right) ,
\end{equation}
with $G \sim G^\prime$, $d(G,G^\prime)=1$.
\end{defn}

Smooth sensitivity serves as a balance between global and local scenarios. Generally, it mitigates the vulnerability associated with local sensitivity and establishes a lower bound compared to global sensitivity. However, the practical implementation is hindered by computational complexity, which is influenced by the specific query function. Smooth sensitivity can be computed by following:

\begin{equation}
    A^{(k)}(G):=\max _{G^\prime \in \mathcal{G}: d(G, G^\prime) \leq k} L S_f(G) 
\end{equation}

\begin{equation}
    \begin{aligned}
    SS_f(G) & =\max_{k=0,1, \ldots, n} e^{-k \beta}\left(\max _{G^\prime: d(G, G^\prime)=k} L S_f(G^\prime)\right) \\
    & =\max _{k=0,1, \ldots, n} e^{-k \beta} A^{(k)}(G) ,
    \end{aligned}
\end{equation}
where $k$ is the maximum distance of $G$ and $G^\prime$.
\subsection{Shortest Path}
For graph analysis tasks, the shortest path is an important metric with numerous applications, including path scheduling, friend matching, and computing betweenness centrality.

\begin{thm}
(Diameter bound \cite{mukwembi2012note}) Let $G$ be a connected graph with $n$ vertices, for diameter $d$, with $d \neq 3,4$, we have:
\begin{equation}\label{equ:diabound1}
    d \leq \frac{3(n-t)}{\Bar{\delta}+1}-1+\frac{3}{\Bar{\delta}+1} ,
\end{equation}
where $t$ is the number of distinct entries of the degree sequence and $\Bar{\delta}$ is the minimum degree of $G$ (for conflict avoidance, we override the normal degree notation $\delta$ to $\Bar{\delta}$ ).
\end{thm}
\begin{thm}
(Diameter bound \cite{mukwembi2012note}) Let $G$ be a connected graph with $n$ vertices, for diameter $d$, with $d = 3, 4$, we have:
    \begin{equation}\label{equ:diabound2}
        d \leq \frac{3(n-t)}{\Bar{\delta}+1}+1+\frac{3}{\Bar{\delta}+1} .
    \end{equation}
\end{thm}

\section{Asymmetric Properties}\label{sec:4}
In this section, we review and analyze the asymmetric properties of previous works \cite{kotsogiannis2020one,takagi2022asymmetric,soria2017individual}. The asymmetric properties include the asymmetric neighborhood and the asymmetric noise.

\subsection{Asymmetric Neighborhood}
The definition of neighborhood is a critical aspect when designing a DP mechanism. It is related to the query function, the granularity of privacy protection, and even implies assumptions about the background knowledge of adversaries. Thus, changing the definition of neighborhood is the primary task or challenge in adapting DP to preserve the privacy of sensitive properties.

The original definition is that neighbors differ in one record. That is, for neighbors $x$ and $y$, $d(x,y)=1$ holds. This definition implies a symmetric relationship between $x$ and $y$.

\begin{lem}\label{symmp}
(Symmetric Property) Let $\mathcal{M}$ be the classic Laplace mechanism of $\varepsilon$-DP. For neighbors $x\sim y$ and any outputs $O \in range(\mathcal{M})$, we have:
\begin{equation}
        \operatorname{Pr}[\mathcal{M}(x) \in O] \leq \exp (\varepsilon) \times \operatorname{Pr}\left[\mathcal{M}\left(y\right) \in O\right] .
\end{equation}
\text{and}
\begin{equation}
        \operatorname{Pr}[\mathcal{M}(y) \in O] \leq \exp (\varepsilon) \times \operatorname{Pr}\left[\mathcal{M}\left(x\right) \in O\right] .
\end{equation}
\end{lem}

Let us assume $x\sim y$ is derived from the following neighborhood definition: neighbors are copies of each other with one record added or removed. With this definition, we can observe: if $y$ is the neighbor of $x$ with one extra record ($x \rightarrow y$), $x$ is also the neighbor of $y$ with one record removed ($y \rightarrow x$). Consequently, the symmetric relationship of $x \leftrightarrow y$ is drawn from the definition of the symmetric neighborhood.

From another point of view, the unbalanced relationship between $x$ and $y$ is also due to the asymmetric environment. The environment limits the range of adjacent data sets. Let us look back at some definitions that have the same asymmetrical characteristic.

\begin{defn}
(Individual Differential Privacy, iDP \cite{soria2017individual}) Given a dataset $x$, we can say that a randomized mechanism $\mathcal{M}$ satisfies $\varepsilon$-iDP at $x$ if, for any neighbor $y$ of $x$ and any output $O \in range(M)$, have:
\begin{equation}
    \begin{gathered}
\exp (-\varepsilon) \times \operatorname{Pr}\left(\mathcal(M)\left(y\right) \in O\right) \leq \operatorname{Pr}(\mathcal{M}(x) \in O) \\
\leq \exp (\varepsilon)\times \operatorname{Pr}\left(\mathcal{M}\left(y\right) \in O\right) .
\end{gathered}
\end{equation}
\end{defn}
The notion of iDP is similar to the lemma \ref{symmp}. The neighbors $x$ and $y$ are not interchangeable. $y$s are all the possible neighbors around $x$, but with the exchange occurring, $x$s are all the possible neighbors of $y$ that the risk of potential privacy leakage cannot be afforded by $\mathcal{M}$. The reason is that, to preserve the utility of the target dataset $x$, the magnitude of the noise in $\mathcal{M}$ is calibrated to the sensitivity of $x$, not $y$.

For the purpose of iDP focus on preserving the privacy of individual, it can not provide the same group privacy guarantees as DP.
\begin{lem}
(Group Differential Privacy) Given that a mechanism $\mathcal{M}$ satisfies $\varepsilon$-DP for neighbors $x,y \in \mathbb{N}^n$ with $d(x,y)=1$. For neighbors $x^\prime,y^\prime \in \mathbb{N}^n$ with $d(x^\prime,y^\prime)=k$, we see that $\mathcal{M}$ satisfies $k\varepsilon$-DP.
\end{lem}
\begin{defn}
(Group Individual Differential Privacy \cite{soria2017individual}) For a dataset $x$, a randomized mechanism $\mathcal{M}$ satisfies $\left(\varepsilon_1,...,\varepsilon_k\right)$-group iDP if, for all $y$s with $d(x,y)=i$ where $i \in [k]$ and $O \in range(M)$, have:
\begin{equation}
    \begin{gathered}
\exp (-\varepsilon_i) \times\operatorname{Pr}\left(\mathcal(M)\left(y\right) \in O\right) \leq \operatorname{Pr}(\mathcal{M}(x) \in O) \\
\leq \exp (\varepsilon_i)\times \operatorname{Pr}\left(\mathcal{M}\left(y\right) \in O\right) .
\end{gathered}
\end{equation}
\end{defn}
Indeed, the group property of iDP is a description of an aggregated form with $i$ iDPs. It achieves the privacy guarantee for $d(x,y)=k$ by treating a group of individuals as a single individual and then applying the iDP. A notable fact is that group iDP can achieve better utility than group in DP, although the group property of iDP is more irregular than other DP variants.

\begin{defn}
(One-sided Differential Privacy, OSDP \cite{kotsogiannis2020one}) Let $P$ be a policy function mapping an individual record $r$ to $0$ or $1$ if $r$ is sensitive or nonsensitive, respectively. Given that the dataset $y$ is $P$-neighbor to $x$, denoted as $x \overset{P}{\rightarrow} y$. Let $\mathcal{M}$ be a randomized algorithm satisfies $(P,\varepsilon)$-OSDP, for any $O \in range(\mathcal{M})$, we have:
\begin{equation}
    \operatorname{Pr}[\mathcal{M}(x) \in O] \leq \exp (\varepsilon) \times \operatorname{Pr}\left[\mathcal{M}\left(y\right) \in O\right] .
\end{equation}
\end{defn}

\begin{defn}
($P$-Neighbors in OSDP) Given databases $x$ and $y$, we call them P-neighbors if, for the policy function $P$, $\forall r\in \{x \backslash (x\cap y)\}$, $r$ is sensitive and $\exists r^\prime \neq r, r^\prime \in \{y \backslash (x\cap y)\}$ is sensitive or nonsensitive.
\end{defn}

The $P$-neighbor is an asymmetric relationship in which $x$ is $P$-neighbor to $y$, but not vice versa; that is, $x \overset{P}{\rightarrow} y$ does not imply $y \overset{P}{\rightarrow} x$. $y$ is derived from $x$ by replacing a sensitive record $r$ in $x$ with a different record $r^\prime$. Note that $r^\prime$ is sensitive or non-sensitive.

The policy function $p$ in OSDP plays a nontrivial role, which decides whether a record $r$ in $x$ is sensitive or not. And OSDP preserves only the privacy of sensitive records, but directly exposes all nonsensitive records. If all the records are sensitive, OSDP provides privacy guarantee for all records as standard DP (the first formal DP definition); and if all records are nonsensitive, no privacy leakage occurs. That is, OSDP can be seen as an extension of DP.

Similarly, a DP variant of \cite{takagi2022asymmetric} was proposed based on a similar policy function and neighborhood.

\begin{defn}
(Asymmetric Differential Privacy, ADP \cite{takagi2022asymmetric}) Let $p$ be a policy function mapping a record $r$ to $\{Flase, True\}$. Given a randomized mechanism $\mathcal{M}$ satisfies $(\varepsilon,p)$-ADP with database $x$ is $p$-neighbor to $y$ (denoted as $x\overset{p}{\sim} y$), if for all outputs $O\in range(M)$, have:
\begin{equation}
        \operatorname{Pr}[\mathcal{M}(x) \in O] \leq \exp (\varepsilon) \times \operatorname{Pr}\left[\mathcal{M}\left(y\right) \in O\right] .
\end{equation}
\end{defn}

\begin{defn}
($p$-Neighbors in ADP) Given databases $x$ and $y$, we say that $x$ is $p$-neighbor to $y$, if (1) $x$ and $y$ differ in records $r$ and $r^\prime$ where $r\in x$ and $r^\prime\in y$; (2) $p(r)=True \text{ and }p(r^\prime)=False $.
\end{defn}

The policy functions of OSDP and ADP play the same role that assign a binary property (i.e., $\{0,1\}$ or $\{True,False\}$) to each record. $p$-neighbors in ADP are also asymmetric, as the operation to construct possible $p$-neighbors is to replace the $True$ record with $Flase$ so that the $p$-neighbors of $y$ exclude $x$.

Leaving policy functions aside, we can observe that the asymmetric neighborhood can meet the requirements of DP as a variant of symmetric with different privacy concerns.

\subsection{Asymmetric Noise}
Before introducing our extension, let us revisit some implementations that adjust with the asymmetric neighborhood. Although the neighborhood of iDP is asymmetric, the concrete mechanism is common (discrete) Laplace mechanisms. Let us jump to OSDP.

\begin{defn}
(One-Sided Laplace in OSDP \cite{kotsogiannis2020one}) Assume $z$ be a random variable drawn from the One-Sided Laplace Distribution (the symmetric version of exponential distribution) with probability density function:
\begin{equation}
    f(z ; \lambda)=\left\{\begin{array}{cc}
\lambda e^{\lambda z} & z \leq 0 \\
0 & z>0 .
\end{array}\right.
\end{equation}
Let $f$ be a count query on the nonsensitive part of the database. $x_{ns}$ (or $y_{ns}$) is the nonsensitive part of the target database $x$ (or the neighboring database $y$). The release of $|x_{ns}|+z$ under the privacy guarantee of OSDP. Here, we have $|x_{ns}| \leq |y_{ns}|$ for a sensitive record in $x$ that may be replaced by a nonsensitive one.
\end{defn}

\begin{defn}\label{def:Alap}
(Asymmetric Laplace Mechanism \cite{takagi2022asymmetric}) Given a query function $f:\mathcal{X}\rightarrow \mathcal{R}$. Assume that $x \in \mathcal{X}$ and $y \in \mathcal{X}$ are $p$-neighbors. Let $\alpha= +1$ if $f(x)\leq f(y)$ or $\alpha= -1$ if $f(x)\geq f(y)$. We have Asymmetric Laplace Mechanism (ALap): $ALap_{\varepsilon,f}(x) = f(x) + z$ satisfies $\varepsilon$-DP where $\varepsilon$ is the privacy parameter and $z$ is a random variable drawn from the following distribution:
\begin{equation}
    f(z ; \alpha)=\left\{\begin{array}{cc}
    \frac{\varepsilon}{GS(f)} \exp \frac{-\operatorname{\alpha}z \varepsilon}{GS(f)} & \left(\operatorname{\alpha}z \geq 0\right) \\
    0 & \left(\operatorname{\alpha}z < 0\right) .
    \end{array}\right.
\end{equation}
\end{defn}

Given a query function $f$, if any $p$-neighbors $x$ and $y$ satisfy monotonicity, that is, $f(x)\leq f(y)$ (monotonically increasing) or $f(x) \geq f(y)$ (monotonically decreasing), we can use the asymmetric Laplace mechanism instead of the symmetric Laplace mechanism. Essentially, the asymmetric Laplace distribution is an exponential distribution or its symmetric form. The DP guarantee can be preserved by ALap since $f$ is monotonic over $p$-neighbors in ADP.

Similarly, the one-sided Laplace has the same route as ALap. The query function $f$ in one-sided Laplace outputs all counts on nonsensitive records. Thus, $f$ increases monotonically with $|x_{ns}| \leq |y_{ns}|$. However, on the other hand, the random variable $z$ is negative, which is drawn from the symmetric exponential distribution. The result, $f(x_{ns}) + z$, is under the privacy guarantee of $\varepsilon$-DP. In addition, how to generalize $f$ is not covered in OSDP since $f$ is just a particular instance.

With the observation of OSDP and ADP, exponential noise (or its symmetric form) can provide the DP guarantee with improved utility if the neighborhood is asymmetric. This thought is reflected in both OSDP and ADP, although the main motivation of OSDP is to obscure the distinguishability of sensitive records from insensitive ones, and the motivation of ADP is to prevent two-sided errors. Thus, we need to refine the definition to clarify its scope of application and tackle the limitations.

\section{Asymmetric Neighborhood Differential Privacy}\label{sec:5}
In this section, we define asymmetric neighbors and propose our asymmetric Laplace mechanism. We then combine smooth sensitivity with our mechanism for enhanced utility and privacy.
\subsection{Global Asymmetry}
\begin{defn}
(Asymmetric Neighbors) Given two databases $x,y \in \mathcal{X}^n$, we call $x$ and $y$ are asymmetric neighbors if $y \in N(x)$ but $x \notin N(y) $, where $N(\cdot)$ is an operation or condition to obtain neighbors, denoted as $x\rightarrow y$.
\end{defn}
The definition of asymmetric neighbors is the abstract form of $p$-neighbors in OSDP and ADP, without considering any policy function. In fact, the policy function is flexible as the application scenarios vary. Essentially, it is a mapping of record statuses. Thus, we remove the policy function to make our definition more explicit.

The symmetric neighbors in standard DP are a strict concern for privacy, that is, considering the worst case of privacy breach. However, conservative concerns can sometimes harm the utility of data caused by excessive privacy preservation. An intuitive thought is that utility can be improved by taking into account the sensitivity of asymmetric neighbors. However, with global sensitivity, the sensitivity of both asymmetric and symmetric neighbors remains the same.
\begin{lem}
The asymmetric and symmetric neighbors have the same global sensitivity.
\end{lem}
Let us define $N(\cdot)$ as adding a record. That is, for any asymmetric neighbors $x, y \in \mathcal{X}^n$, we have $x\rightarrow y$ where $y$ has an additional record $r$ than $x$. This neighborhood relationship illustrates our target for preserving privacy: \textit{ for every additional record $r$}. Since adversaries cannot break the indistinguishability between the query outputs of $x$ and $y$, the privacy of the record $r$ has been preserved. Symmetrically, if $y \in N^\prime(x)$ but $y$ is one record $r$ less than $x$, the target is all the actual records in $x$. When removing any $r\in x$, the indistinguishability will still be preserved, so adversaries cannot learn about the existing records in $x$. The different targets reflect the difference between $N(\cdot)$ and $N^\prime(\cdot)$.

The privacy targets mentioned above are distinct, but both possess the same level of global sensitivity. For neighbors $x$ and $y$ are from the universe $\mathcal{X}^n$, $x\rightarrow y$ and $y\rightarrow x$ can be achieved by $N(\cdot)$ and $N^\prime(\cdot)$, respectively. Since any extra record in $y$ (from the view of $N(\cdot)$) can be the existing record in $y$ (from the view of $N^\prime(\cdot)$), $x\rightarrow y$ and $y\rightarrow x$ share the same global sensitivity. Therefore, asymmetric neighbors have the same global sensitivity, which is also the global sensitivity in symmetric neighbors. More concretely, with the example of counting, whatever the $N(\cdot)$ is, adding or removing a record, the global sensitivity is $1$. And for triangle counting, the global sensitivity is $n-2$ for adding or removing an edge.

\begin{defn}
($f$-Asymmetric Sensitivity) Let $f: \mathcal{X}^n \rightarrow \mathcal{R}^d$ be a query function. With $N(\cdot)$, for any $y\in N(x)$ if given any $x \in \mathcal{X}^n$, we have $f$-asymmetric sensitivity:
\begin{equation}
    AS_f :=\sup _{x \underset{x,y\in \mathcal{X}^n}{\rightarrow}  y}\left\|f(x)-f\left(y\right)\right\|_1 .
\end{equation}
\end{defn}
$AS_f$ is a global sensitivity over query function $f$ which measures the maximum difference of the outputs of $f$ over $x$ and $y$. With the asymmetric neighbors, we can easily obtain the following result for some query functions.
\begin{lem}
(Monotonic Property of $f$) For any $x\rightarrow y$, $f$ is monotonically decreasing (or increasing) if $f(x)_i\geq f(y)_i$ (or $f(x)_i\leq f(y)_i$ ) for any $i \in [d]$.
\end{lem}
The ADP has a similar definition for the monotonic property; however, it is associated with $p$-sensitivity, which may be confusing as to the source of this monotonicity. Actually, monotonicity is related to the query function $f$ over neighbors $x\rightarrow y$, especially for some monotonic query functions, such as counting or summing.
\begin{defn}
(Global Asymmetric Differential Privacy) Given any neighbors $x,y \in \mathcal{X}^n$ over operation $N(\cdot)$,that is, $y \in N(x)$, a randomized algorithm $\mathcal{M}$ satisfies $\varepsilon$-global Asymmetric Differential Privacy ($\varepsilon$-gADP) if, for any outputs $O \in range(\mathcal{M})$, we have:
\begin{equation}
    \operatorname{Pr}[\mathcal{M}(x) \in O] \leq \exp (\varepsilon) \times \operatorname{Pr}\left[\mathcal{M}\left(y\right) \in O\right] .
\end{equation}
\end{defn}

The term \textit{global} is from global neighbors, which is to be distinguished from Subsection \ref{subsec:iasy}. Unlike standard DP, gADP considers the indistinguishability of $x$ and $y$ under asymmetric operation $N(\cdot)$. The standard DP can offer a higher level of privacy assurance for symmetric neighbors than gADP does. Thus, we have the following observation.
\begin{lem}
If a randomized algorithm $\mathcal{M}$ satisfies DP, it also satisfies gADP.
\end{lem}

\subsection{Individual Asymmetry}\label{subsec:iasy}
With the same concern for iDP, the standard DP (or gADP) provides a more strict privacy guarantee than what we need intuitively for our database. We do not need to take into account all the potential neighboring universes if we want to answer queries in a confidential manner as data holders. Thus, it is sufficient to maintain only indistinguishability between the actual database $x$ and its neighbors.

\begin{defn}
($f$-Individual Asymmetric Sensitivity) Let $f: \mathcal{X}^n\rightarrow \mathcal{R}^d$ be a query function. With $N(\cdot)$, given the actual database $x\in \mathcal{X}$ for any $y \in N(x)$, we have $f$-individual asymmetric sensitivity:
\begin{equation}
    iAS_f:=\sup _{x \underset{y\in \mathcal{X}^n}{\rightarrow} y}\left\|f(x)-f\left(y\right)\right\|_1 .
\end{equation}
\end{defn}

$iAS_f$ is similar to $LS_f$ but with asymmetric neighbors. The utility of the data can be improved because $AS_f$ is the worst case of $iAS_f$. In most cases, $iAS_f$ is smaller than $AS_f$.
\begin{defn}
(Individual Asymmetric Laplace Mechanism) Let $f: \mathcal{X}^n \rightarrow \mathcal{R}^d$ be a query function. Given $x\rightarrow y$ over operation $N(\cdot)$ for actual database $x\in \mathcal{X}^n$ and all $y\in N(x)$, we have a randomized algorithm $\mathcal{M}$ is an Individual Asymmetric Laplace Mechanism (IALap) that satisfies $\varepsilon$-Individual Asymmetric Differential Privacy (or $\varepsilon$-IADP) for all $O \in range(\mathcal{M})$, if:
\begin{equation}
    \operatorname{Pr}[\mathcal{M}(x) \in O] \leq \exp (\varepsilon) \times \operatorname{Pr}\left[\mathcal{M}\left(y\right) \in O\right] ,
\end{equation}
where $\mathcal{M}(x)=f(x)+(z_1,...,z_d)$, and $(z_1,...,z_d)$ are independent random variables drawn from exponential distribution, denoted as $Exp^+(\lambda)$:
\begin{equation}
    Exp^+(\lambda) =\left\{
    \begin{array}{cc}
        0 & (z < 0) \\
        \lambda \exp {\left(-z\lambda\right)} & (z \geq 0 ) ,
    \end{array}\right.
\end{equation}
if $f$ is monotonically decreasing over $x\rightarrow y$, or from the symmetric version of exponential distribution, denoted as $Exp^-(\lambda)$:

\begin{equation}
    Exp^-(\lambda) =\left\{
    \begin{array}{cc}
        \lambda \exp {\left(z\lambda\right)} & (z \leq 0) \\
        0 & (z > 0 ) ,
    \end{array}\right.
\end{equation}
if $f$ is monotonically increasing over $x\rightarrow y$, where $\lambda=\frac{\varepsilon}{iAS_f}$.
\end{defn}

However, with the concerns of \cite{nissim2007smooth}, $iAS_f$ (or $LS_f$) may be at risk in terms of privacy breaches. Consider two neighbors $D_1=\{0,0,0,0,1\}$ and $D_2=\{0,0,0,1,1\}$. Given a query function $f_{med}$ that returns the median. We have $f_{med}(D_1)=f_{med}(D_2)=0$. However, for $LS_f$, there exist $L S_{f_{m e d}}(D_1)=0$ and $L S_{f_{m e d}}(D_2)=1$. That is, the median of $D_1$ will be answered without masking. And the indistinguishability of $D_1$ and $D_2$ is destroyed unless we tolerate an approximation factor $\delta$ since the probability of answering $0$ is very different.
\begin{lem}
For $D_1$ and $D_2$, with a discrete Laplace mechanism, local sensitivity will compromise privacy if $\varepsilon < \ln{\left(1 + \sqrt{2}\right)}$.
\end{lem}
$Proof$: The discrete Laplace mechanism $\mathcal{M} = f(D) + z$ where $z$ is a random variable drawn from:
\begin{equation}
    \operatorname{Pr}(z=i)=\frac{1-\alpha}{1+\alpha} \alpha^{|i|} .
\end{equation}
If $\alpha = \exp \left(-\varepsilon / L S_f(D)\right)$, $\mathcal{M}$ satisfies $\varepsilon$-iDP \cite{soria2017individual}. However, for $S=\{0\}$, we have:
\begin{equation}
\begin{aligned}
    \operatorname{Pr}[\mathcal{M}(D_1)\in S] &\leq \exp(\varepsilon) \times \operatorname{Pr}[\mathcal{M}(D_2)\in S] \\
    1 &\leq \exp{(\varepsilon)}\times \frac{1-\exp{(-\varepsilon)}}{1+\exp{(-\varepsilon)}}\\
    \varepsilon &\geq \ln{\left(1 + \sqrt{2}\right)} .
    \end{aligned}
\end{equation}
$\hfill\blacksquare$

The $\varepsilon$-indistinguishability of $D_1$ and $D_2$ can be maintained if $\varepsilon$ is greater than or equal to $\ln{\left(1 + \sqrt{2}\right)}$, which is contrary to the expectation that the smaller the $\varepsilon$, the better the preservation of privacy. For asymmetric neighbors, the paradox also exists. Thus, we should avoid using local sensitivity (or $f$-individual asymmetric sensitivity) to prevent potential privacy breach issues.

Smooth sensitivity is a suitable sensitivity that is a minimum smooth upper bound on $f$-individual asymmetric sensitivity. It avoids privacy leakage issues while providing greater utility compared to global sensitivity. With the maximum number of records that differ by less than or equal to $1$, we can get smooth sensitivity $SS_f$ by:
\begin{lem}
For query function $f$ and asymmetric neighbors $x,y\in \mathcal{X}^n$ over operation $N(\cdot)$ where $d(x,y)\leq 1$, we have:
\begin{equation}
\small{ SS_f = \max\left\{ iAS_f(x),\max_{\underset{y\in N(x)}{d(x,y)=1}}{\left(iAS_f(y) \right)}\times \exp{(-\beta)}  \right\}} .
\end{equation}
\end{lem}

To provide a DP guarantee, the noise calibrated to $SS_f$ should be drawn from the admissible noise distribution with parameters $\alpha$ and $\beta$ \cite{nissim2007smooth}. Laplace distribution is an admissible distribution; thus, for the exponential distribution, a similar result also holds.

\begin{lem}
For $\delta \in(0,1)$, exponential distribution (or its symmetrical version) is $(\alpha, \beta)$-admissible with $\alpha=\frac{\varepsilon}{2}$ and $\beta=\frac{\varepsilon}{2 \ln (2 / \delta)}$.
\end{lem}
\begin{lem}
With smooth sensitivity, IALap can provide $(\varepsilon, \delta)$-IADP guarantee for $\delta \in(0,1)$.
\end{lem}
The proofs follow the same routes as in \cite{nissim2007smooth}.

\section{Privately Distance Query}\label{sec:6}
In this section, we study the problem of how to answer distance queries in graphs privately. With the concern of utility, we separately discuss two asymmetric neighborhood settings: adding an edge and removing an edge. In each setting, we provide ways to preserve the privacy of distances.
\subsection{Problem}
For simplicity, let $G=(V,E)$ be a simple, connected, and unweighted graph. The 'simple' implies that the graph is undirected and does not have loops or multiple edges. The 'connected' means that there exists a reachable path between any vertices. And the 'unweighted' means that we do not consider the weights on the edges. Let $f_{dis,G}:\left(u,v\right) \rightarrow \mathcal{R}$ be a distance query function that outputs the distance of the pair of vertices $(u,v)$. Here, we use $e(u,v)$ to denote the edge between the vertex $u$ and $v$.

Edges represent relationships between vertices. For example, the edge $e(u,v)$ in the social graph reflects the relationship of user $u$ and $v$. Note that the relationship can be more finely grained in the weighted graph, e.g., common friends, close friends, family members, lovers, etc. But in our assumptions about the unweighted graph, we uniformly consider it as a 'relationship'. With the ethical and legal requirements for privacy, users $u$ and $v$ have the right to refuse to expose their relationship when data must be disclosed to third parties. More concretely, if the third party queries the distance of $u$ and $v$, how to perturb the query results so that the relationship between $u$ and $v$ remains private is the issue that the data owner must solve.

DP-related techniques are helpful to address this issue; however, the global sensitivity of $f_{dis,G}$ will damage the utility of the results if we apply the classic Laplace mechanism. Let us review the global sensitivity of $f_{dis,G}$ in symmetric neighbors.
\begin{lem}
    For a graph $G$ with $n$ vertices, $GS_{f_{dis,G}} = \infty$.
\end{lem}
Since the operation $N(\cdot)$ is adding or removing an edge from the graph $G$, removing will cause vertices $u$ and $v$ to be disconnected. Unfortunately, it is not possible to obscure the $\infty$ outcomes with regular random noise, since we cannot set $\varepsilon$ to $\infty$. Also, refusing to answer this kind of results will compromise privacy. Even if we do not consider this extreme case, that is, we assume that $G$ is $2$-connected, we still have $GS_{f_{dis,G}}=O\left(n\right)$. The $G$ is $2$-connected means that any pair of vertices $u$ and $v$ in $G$ are kept connected if any edge is removed from $G$. In the real world, the graphs are usually large and complex, and thus $O\left(n\right)$ is also intolerable.

To improve the utility, we can consider the asymmetric neighborhood, that is, $N(\cdot)$ refers to adding an edge and $N^\prime(\cdot)$ is removing an edge.
\begin{lem}
$GS_{f_{dis,G}}$ is monotonically decreasing (or increasing) over $N(\cdot)$ (or $N^\prime(\cdot)$).
\end{lem}

Intuitively, adding an edge $e(u,v)$ for any $u,v\in V$ can decrease $f_{dis,G}\left(u,v\right)$ to $1$; and removing an edge $e(u,v)$ for any $u,v\in V$ will increase $f_{dis,G}\left(u,v\right)$ from $1$ to at most $n-1$ (or $\infty$ if disconnected). And for $s,t\in V\backslash\{u,v\}$, $f_{dis,G}\left(s,t\right)$ may remain the same, or decrease for adding and increase for removing, respectively. Thus, $f_{dis,G}$ is a monotonic query function in different asymmetric neighborhoods.
\begin{figure}
    \centering
    \includegraphics[width=0.7\linewidth]{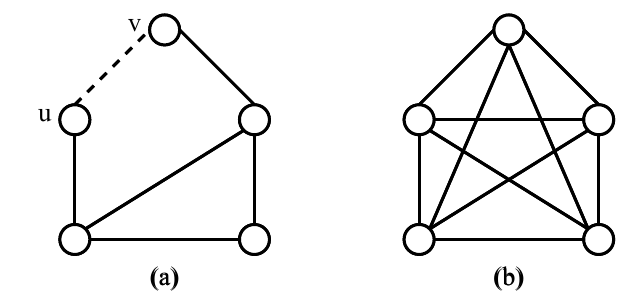}
    \caption{Add an edge.}
    \label{fig:add}
\end{figure}
\subsection{Adding an Edge}
Let us begin with considering $N(\cdot)$ to be adding an edge. In this setting, $f_{dis,G}$ is monotonically decreasing. Thus, we can privately answer $f_{dis,G}$ using IALap. Before applying IALap, we first need to calculate its sensitivity.
\begin{thm}
Given the actual graph $G=\left(V,E\right)$, for the diameter function $dia(\cdot)$, we have:
\begin{equation}
\begin{aligned}
    iAS_{f_{dis,G}} &= \max_{u,v \in V}\lVert f_{dis,G}(u,v) - f_{dis,G^\prime}(u,v) \rVert_1 \\
    & = dia(G) - 1\\
    & =\max_{u,v\in V} f_{dis,G}\left(u,v\right) -1 .
    \end{aligned}
\end{equation}
\end{thm}
When adding an edge to $G$, the distance for any pair $(u,v)$ cannot be larger than before, as the edge will only make the shortest path unchanged or shorter. The diameter represents the maximum distance in any connected graph. Thus, we can obtain $iAS_{f_{dis,G}}$ if the newly added edge joins exactly two endpoints of the diameter. In the asymptotic analysis, the complexity of $iAS_{f_{dis,G}}$ is $O\left(n\right)$ since the maximum diameter is $n-1$. For theoretical analysis, scaling the upper bound of the diameter is feasible.
\begin{lem}
With Equation \ref{equ:diabound1} and \ref{equ:diabound2}, we have:
\begin{equation}
    iAS_{f_{dis,G}} \leq \frac{3(n-t)}{\Bar{\delta}+1}+O(1) ,
\end{equation}
where $t$ is the irregularity index of $G$ and $\Bar{delta}$ is the minimum degree of $G$.
\end{lem}

As demonstrated in Fig. \ref{fig:add}, Fig. \ref{fig:add}(a) is a connected graph and  Fig. \ref{fig:add}(b) is a complete graph. The dashed line is the added edge $e(u,v)$. Before adding $e(u,v)$,
$f_{dis,G}\left(u,v\right)=3$. However, after adding $e(u,v)$, $f_{dis,G}\left(u,v\right)=1$, the maximum change occurs.

However, if the actual graph is a complete graph, as shown in Fig. \ref{fig:add}(b), $iAS_{f_{dis,G}}$ is $0$. At this point, the results of the query are disclosed without masking, leading to privacy breaches. To address this issue, a reasonable approach would be to fix $iAS_{f_{dis,G}}$ to $1$ since each edge contributes equally to the distance query with a value of $1$.

\begin{lem}
For $f_{dis,G}$ and for all $G^\prime \in N(G)$, we have:
\begin{equation}
    \begin{aligned}
        SS_{f_{dis,G}} &= iAS_{f_{dis,G}} .
    \end{aligned}
\end{equation}
\end{lem}
$Proof$: Since $f_{dis,G}$ is monotonically decreasing for all $G^\prime \in N(G)$, we have $dia(G) \geq dia(G^\prime)$. And with $\exp{\left(-\beta \right)} < 1$, the lemma holds.
$\hfill\blacksquare $

\begin{algorithm}
  \caption{Answer distance with positive noise}
  \label{alg:1}
  \begin{algorithmic}[1]
  \renewcommand{\algorithmicrequire}{\textbf{Input:}}
  \renewcommand{\algorithmicensure}{\textbf{Output:}}
  \REQUIRE Actual graph $G=(V,E)$, query target $u,v$, privacy parameters $ \varepsilon, \delta$, noise parameters $\alpha=\frac{\varepsilon}{2}, \beta=\frac{\varepsilon}{2 \ln (2 / \delta)}$
  \ENSURE  $\hat{f}_{dis,G}(u,v)$
    \STATE $dia(G)\leftarrow$ calculate diameter of $G$
   \IF {$dia(G) = 1$}
   \STATE $iAS_{f_{dis,G}} = 1$
   \ELSE
   \STATE $iAS_{f_{dis,G}} = dia(G)- 1$
   \ENDIF
   \STATE $SS_{f_{dis,G}} \leftarrow iAS_{f_{dis,G}}$
   \STATE $f_{dis,G}(u,v)\leftarrow$ calculate the distance between $(u,v)$
   \STATE $\hat{f}_{dis,G}(u,v)\leftarrow f_{dis,G}(u,v)+\frac{SS_{f_{dis,G}}}{\alpha}\times Exp^+(1)$
   \STATE $\hat{f}_{dis,G}(u,v)\leftarrow \hat{f}_{dis,G}(u,v) - \frac{SS_{f_{dis,G}}}{\alpha}\times \ln{2} $
   \STATE $\hat{f}_{dis,G}(u,v) \leftarrow $ random rounding $\hat{f}_{dis,G}(u,v)$
   \IF {$\hat{f}_{dis,G}(u,v) > n-1$}
        \STATE $\hat{f}_{dis,G}(u,v) \leftarrow n-1$
   \ENDIF
  \RETURN $\hat{f}_{dis,G}(u,v)$
  \end{algorithmic}
\end{algorithm}

The Alg. \ref{alg:1} demonstrates the process of privately answering the distance of $(u,v)$. First, we calculate the diameter of the actual graph $G$, denoted as $dia(G)$. If $dia(G)=1$, then $G$ is a complete graph. Thus, we set $iAS_{f_{dis,G}}=1$. But if $G$ is not complete, we have $iAS_{f_{dis,G}}=dia(G) - 1$. Then, we have $SS_{f_{dis,G}} = iAS_{f_{dis,G}}$. Since $Exp^+(1)$ is not unbiased, we subtract its median value $\ln{2}$ from the query result to reduce the overall error. To follow the integer property of distance, we round the answer with random rounding.

\begin{lem}
Random rounding will not cause an additional mistake when estimating the expectation.
\end{lem}
$Proof$: Given a random variable $z$, let $a$ and $b$ be the integer and fractional parts of $z$, respectively, such that $z = a+ b$. The random rounding result is denoted by $\Bar{z}$.
\begin{equation}
    \Bar{z}= \begin{cases}a+1 & \text { w.p. } b \\ a & \text { w.p. } 1-b .\end{cases}
\end{equation}
\begin{equation}
    \mathbb{E}\left[\Bar{z}\right]=\left(a+1\right) \times b+a \times\left(1-b\right)=a+b .
\end{equation}
$\hfill\blacksquare $

After random rounding, we truncate the answer if it is greater than $n-1$ since $n-1$ is the maximum distance in any graph with $n$ vertices. Random rounding and truncation operations can also be regarded as post-processing steps. Finally, the answer $f_{dis,G}(u,v)$ can be answered with $(\varepsilon,\delta)$-IADP by adding noise $\frac{SS_{f_{dis,G}}}{\alpha}\times Lap^{+}(1)$. Fortunately, since $SS_{f_{dis,G}} = iAS_{f_{dis,G}}$, the answer is masked by $\varepsilon^\prime$-IADP guarantee where $\varepsilon^\prime = \frac{\varepsilon}{2}$.

\subsection{Removing an Edge}
Different with adding an edge, removing an edge requires stronger assumptions to make $G$ always connected, even after removing an edge. Assuming that $G$ is an $3$-edge connected graph, we can avoid $iAS_{f_{dis,G^\prime}}=\infty$ for all $G^\prime \in N^\prime(G)$ where $N^\prime(\cdot)$ refers to removing an edge.

To privately answer the query, we also need to calculate the sensitivity. A naive way is to calculate $iAS_{f_{dis,G}}$ and $iAS_{f_{dis,G^\prime}}$ for all $G^\prime \in N^\prime(G)$. However, we can observe that the complexity of the calculation is up to $O\left( {\lvert E\rvert}^2 n^2 \right)$. For ${\lvert E\rvert} \geq \left\lceil{\frac{kn}{2}}\right\rceil$, where $k$ is the smallest number of edges removed that makes $G$ unconnected, the complexity is $\Omega\left(n^4 \right)$. For large graphs, the costs are not affordable.

\begin{defn}
($t$-Shortest Path of $\left(u,v\right)$) Given $t \in \mathbb{N}$ and a connected graph $G$ with at least $2$ vertices, for the vertex pair $(u,v)$, $t$-shortest path of $(u,v)$,an edge sequence, denoted as $P^t(u,v)$, is the shortest path of $(u,v)$ in $G\backslash \{ P^1(u,v), p^2(u,v), ... ,p^{t-1}(u,v) \}$ where all edges passed by $\{ P^1(u,v), p^2(u,v), ..., p^{t-1}(u,v) \}$ have been removed.
\end{defn}

$P^t(u,v)$ describes a kind of shortest paths. $P^1(u,v)$ is the shortest path of $(u,v)$ in $G$. And then, $P^2(u,v)$ is the shortest path of $(u,v)$ in $G\backslash \{P^1{u,v} \}$ where the edges passed by $P^1(u,v)$ have been removed. That is, $P^2(u,v)$ is the second shortest path that does not interact with $P^1(u,v)$. With $t$ increasing, we have:
\begin{lem}
For any vertex pair $(u,v)$ in $G$ and possible integer $t > 1$, $P^t(u,v) \bigcap P^{t-1}(u,v) = \varnothing$.
\end{lem}

Note that there may exist two or more shortest paths with the same lengths of $(u,v)$ in $G$, thus $P^1(u,v)$ and $P^2(u,v)$ can have the same lengths.

With the definition of $t$-shortest path, we can calculate the sensitivity:
\begin{thm}
Given the actual graph $G=\left(V,E\right)$, for all $G^\prime \in N^\prime(G)$, we have:
\begin{equation}
    \begin{aligned}
        iAS_{f_{dis,G}} &=  \max_{\underset{G}{u,v \in V} }\left\lVert f_{dis,G}(u,v) - f_{dis,G^\prime}(u,v) \right\rVert_1 \\
        &= \max_{\underset{G}{u,v \in V} }\left(\left\lvert P^2(u,v)\right\rvert - \left\lvert P^1(u,v)\right\rvert \right) .
    \end{aligned}
\end{equation}
\end{thm}

Note that $P^1(u,v)$ and $P^2(u,v)$ are collections of edges. Then, $iAS_{f_{dis,G}}$ is equal to the maximum difference between $\lvert P^2(u,v)\rvert$ and $\lvert P^1(u,v)\rvert$ where $|\cdot|$ returns the size of the collection. For $G^\prime$, we can get the following corollary.
\begin{cor}\label{cor:1}
Given the actual graph $G=\left(V,E\right)$, for any $G^\prime \in N^\prime(G)$, $e(s,t)$ is the removed edge, we have:
\begin{equation}
    \begin{aligned}
    T &= \left(\left\lvert P^3(s,t)\right\rvert - \left\lvert P^2(s,t)\right\rvert \right) \\
    iAS_{f_{dis,G^\prime}} &= \max_{\underset{G^\prime}{u,v \in V} }\left(\left\lvert P^2(u,v)\right\rvert - \left\lvert P^1(u,v)\right\rvert \right) \\
        &= \max_{\underset{G}{u,v \in V} } \left\{\left(\left\lvert P^2(u,v)\right\rvert - \left\lvert P^1(u,v)\right\rvert\right), T\right\} .
    \end{aligned}
\end{equation}
\end{cor}

Since it is difficult to tell if $iAS_{f_{dis,G}}$ is greater than $iAS_{f_{dis,G^\prime}}$, thus we have to iterate all $G^\prime$s to calculate $SS_{f_{dis,G}}$.
\begin{lem}
For $f_{dis,G}$ and for all $G^\prime \in N^\prime(G)$, we have:
\begin{equation}
    \begin{aligned}
        \phi &= \max_{\underset{G}{u,v \in V} }\left(\left\lvert P^2(u,v)\right\rvert - \left\lvert P^1(u,v)\right\rvert\right) \\
        \psi &= \max_{\underset{G}{s,t \in E} }\left(\left\lvert P^3(s,t)\right\rvert - \left\lvert P^2(s,t)\right\rvert \right) \\
        SS_{f_{dis,G}} &=\max \left\{ \phi, \psi\times \exp\left(-\beta\right) \right\} .
    \end{aligned}
\end{equation}
\end{lem}
$Proof$: With Corollary \ref{cor:1}, we have:
\begin{equation}
    \begin{aligned}
    \max_{G^\prime \in N^\prime(G)}iAS_{f_{dis,G^\prime}} &= \max\left\{\phi,\psi \right\}\\
        \phi &= \max_{\underset{G}{u,v \in V} }\left(\left\lvert P^2(u,v)\right\rvert - \left\lvert P^1(u,v)\right\rvert\right) \\
        \psi &= \max_{\underset{G}{s,t \in E} }\left(\left\lvert P^3(s,t)\right\rvert - \left\lvert P^2(s,t)\right\rvert \right) .
    \end{aligned}
\end{equation}
Since $\exp{\left(-\beta\right)} < 1$, we have $iAS_{f_{dis,G}} > \phi \times \exp{\left(-\beta\right)}$, thus
\begin{equation}
    \begin{aligned}
        SS_{f_{dis,G}} & = \max\left\{iAS_{f_{dis,G}}, \psi\times \exp\left(-\beta\right) \right\} \\
        &= \max \left\{ \phi, \psi\times \exp\left(-\beta\right) \right\} .
    \end{aligned}
\end{equation}
$\hfill\blacksquare $

\begin{algorithm}
  \caption{Answer distance with negative noise}
  \label{alg:2}
  \begin{algorithmic}[1]
  \renewcommand{\algorithmicrequire}{\textbf{Input:}}
  \renewcommand{\algorithmicensure}{\textbf{Output:}}
  \REQUIRE Actual graph $G=(V,E)$, query target $\Bar{u},\Bar{v}$, privacy parameters $ \varepsilon, \delta$, noise parameters $\alpha=\frac{\varepsilon}{2}, \beta=\frac{\varepsilon}{2 \ln (2 / \delta)}$
  \ENSURE  $\hat{f}_{dis,G}(u,v)$
    \STATE $\Phi \leftarrow \varnothing$
    \STATE $\Psi \leftarrow \varnothing$
   \FOR {$u$ in $V$}
       \FOR {$v$ in $V$}
           \IF{$e(u,v) \in E$}
                \STATE $P^2(u,v) \leftarrow$ start execute BFS starting at $u$ in $G\backslash\{e(u,v)\}$
                \STATE $P^3(u,v) \leftarrow$ start execute BFS starting at $u$ in $G\backslash\{P^2(u,v), e(u,v)\}$
                \STATE $\Psi \leftarrow \Psi \cup \left\{\left(\left\lvert P^3(u,v)\right\rvert - \left\lvert P^2(u,v)\right\rvert\right)\right\}$
                \STATE $\Phi \leftarrow \Phi \cup \left\{\left(\left\lvert P^2(u,v)\right\rvert - 1\right)\right\}$
            \ELSE
                \STATE $P^1(u,v) \leftarrow$ start execute BFS starting at $u$ in $G$
                \STATE $P^2(u,v) \leftarrow$ start execute BFS starting at $u$ in $G\backslash\{P^1(u,v)\}$
                \STATE $\Phi \leftarrow \Phi \cup \left\{\left(\left\lvert P^2(u,v)\right\rvert - \left\lvert P^1(u,v)\right\rvert\right)\right\}$
           \ENDIF
       \ENDFOR
   \ENDFOR
   \STATE $SS_{f_{dis,G}} \leftarrow \max\left\{\max\left(\Phi\right), \exp{\left(-\beta\right)} \times \max\left(\Psi\right) \right\}$
   \STATE $f_{dis,G}(u,v)\leftarrow$ calculate the distance between $(u,v)$
   \STATE $\hat{f}_{dis,G}(u,v)\leftarrow f_{dis,G}(u,v)+\frac{SS_{f_{dis,G}}}{\alpha}\times Exp^-(1)$
   \STATE $\hat{f}_{dis,G}(u,v)\leftarrow \hat{f}_{dis,G}(u,v) + \frac{SS_{f_{dis,G}}}{\alpha}\times \ln{2} $
   \STATE $\hat{f}_{dis,G}(u,v) \leftarrow $ random rounding $\hat{f}_{dis,G}(u,v)$
   \IF {$\hat{f}_{dis,G}(u,v) < 1$}
        \STATE $\hat{f}_{dis,G}(u,v) \leftarrow 1$
   \ENDIF
  \RETURN $\hat{f}_{dis,G}(u,v)$
  \end{algorithmic}
\end{algorithm}
With the observation that $G$ is an unweighted graph, we can use Breadth-first search (BFS) to find the shortest path between $u$ and $v$. The worst-case performance of BFS is $O(|V|+|E|)$. Therefore, the complexity of calculating $SS_{f_{dis,G}}$ is $O\left(n^2(V+E) \right)$, representing a substantial reduction compared to the naive form.

Alg. \ref{alg:2} outlines the process of answering the distance with negative noise, with a focus on calculating the smoothness sensitivity. We iterate through all pairs of vertex $(u,v)$ and run BFS from $u$ to find the $t$-shortest path $P^t$ of $(u,v)$. Specifically, if having an edge connect $u$ and $v$, we search for $P^2(u,v)$ and $P^3(u,v)$ and put the length difference on $\Psi$ and $\Phi$. Otherwise we search for $P^1(u,v)$ and $P^2(u,v)$ and put the length difference to $\Phi$. Once the iteration is complete, we can acquire a smooth sensitivity and add a symmetric version of exponential noise calibrated to the smooth sensitivity. We can obtain the response after post-processing includes adding the median value $\ln{2}$, random rounding, and truncation.

The noised answer $\hat{f}_{dis,G}(u,v)$ is under the privacy preservation of $(\varepsilon,\delta)$-IADP by adding noise $\frac{SS_{f_{dis,G}}}{\alpha}\times Lap^{-}(1)$.

\section{Experiment}\label{sec:7}

In this section, we assess the performance of our two proposed algorithms using three real-world datasets and three synthetic datasets. We keep $\delta$ fixed and increase $\varepsilon$ to validate the effectiveness of our algorithms.
\subsection{Datasets}
Since we have two asymmetric neighbor operations $N(\cdot)$ and $N^\prime(\cdot)$, adding an edge and removing an edge, we have to provide two different experiment settings, respectively. The first concern is the choice of datasets. For adding an edge, connected graphs are common in the real world, such as social graphs, road graphs, and citation graphs. Thus, we choose three real-world datasets: EIES, BOTC and TDE, described as follows. Table. \ref{tab:ds} presents a summary of all the dataset statistics.

EIES \cite{freeman1979networkers} is Freeman’s EIES network at time $2$ that contains researchers working on the analysis of social networks and their relationships. The origin dataset is weighted and directed. We clean it by replacing directed edges with undirected edges, ignoring weight information, and then obtain a small graph with $34$ vertices and $474$ edges.

BOTC \cite{kumar2018rev2} it is a trust graph of individuals who engage in trading with Bitcoin on a platform known as Bitcoin OTC. This graph is also directed and weighted. We follow a similar clean process: replacing directed edges by undirected ones, ignoring weight information, and removing isolated vertices. As a result, the actual number of edges is reduced to $192,899$.

TDE \cite{rozemberczki2021multi} is a Twitch social graph of gamers who stream in German. Vertices are the users, and edges are mutual relationships between them. This graph is undirected and weighted. We just need to ignore the related weight information to get a clean graph.

For removing an edge, with the additional graph connectivity requirements of $N^\prime(\cdot)$, we construct Harary graphs \cite{harary1962maximum} to validate our Alg. \ref{alg:2}. The Harary graph $H_{k,n}$ is an example of a $k$-connected graph with $n$ vertices. We construct three Harary graphs, namely $H_{3,200},H_{3,1000},H_{3,5000}$, corresponding to the number of vertices $200$, $1000$, and $5000$, respectively. For simplicity, we use SHG (small Harary graph), MHG (medium Harary graph), and LHG (large Harary graph) to represent $H_{3,200},H_{3,1000}$, and $H_{3,5000}$, respectively.
\begin{table}
  \caption{Datasets Statistics}
  \label{tab:ds}
  \begin{tabular}{ccccc}
    \toprule
    Datasets&Vertices&Edges&Diameter&Average Distance\\
    \midrule
    EIES & $34$ & $474$&$2$&$1.16$\\
    BOTC & $5,881$ & $21,492$&$9$&$3.57$\\
    TDE & $9,498$ & $192,899$&$3$&$2.02$\\
    $H_{3,200}$ &$200$& $740$&$35$&$16.82$\\
    $H_{3,1000}$ &$1,000$& $3,740$&$165$&$79.59$\\
    $H_{3,5000}$ &$5,000$& $9,250$&$815$&$393.35$\\
  \bottomrule
\end{tabular}
\end{table}
\subsection{Parameters}
To measure the utility of IADP and compare it with other solutions, we use all-pair distance average relative error as the core metric. For clarity, we refer to it as the error.
\begin{defn}
(All-pair Distance Mean Relative Error) Let us assume that $D_{u,v}$ refers to the true distance of the pair $(u,v)$. And let $\Bar{D}_{u,v}$ be the distance after adding noise. We have the all-pair distance MRE $\eta$
\begin{equation}
\eta =\frac{1}{n^2-n} \sum_{u,v \in V}\left(\frac{\lvert \Bar{D}_{u,v} -  D_{u,v} \rvert}{D_{u,v}}  \right) ,
\end{equation}
where $V$ is the vertices set and $n = \lvert V \rvert$.
\end{defn}

For privacy parameters $\varepsilon$ and $\delta$, we fix $\delta$ and analyze the effect of $\varepsilon$ on errors. Since $\delta$ is the probability of privacy leakage, there will be a privacy leak of $\delta n$ records for a database with $n$ records empirically. For privacy concerns, we fix $\delta = \frac{1}{10n}$.

Since few works follow our assumption, which involves differentially private distance queries on unweighted graphs, we compare our solution IADP (smoothness sensitivity-based implementation) with Standard Differential Privacy (SDP) and ADP. SDP refers to adding Laplace noise calibrated to the global sensitivity $n-1$ to the distance query result. We have slightly modified the ADP to add $Exp^+$ when adding an edge and $Exp^-$ when removing an edge. Both solutions follow the same post-processing steps, including random rounding and truncation. Additionally, for ADP, we adjust its median value to align with our IADP solution.

\begin{figure*}
    \centering
    \includegraphics[width=1\linewidth]{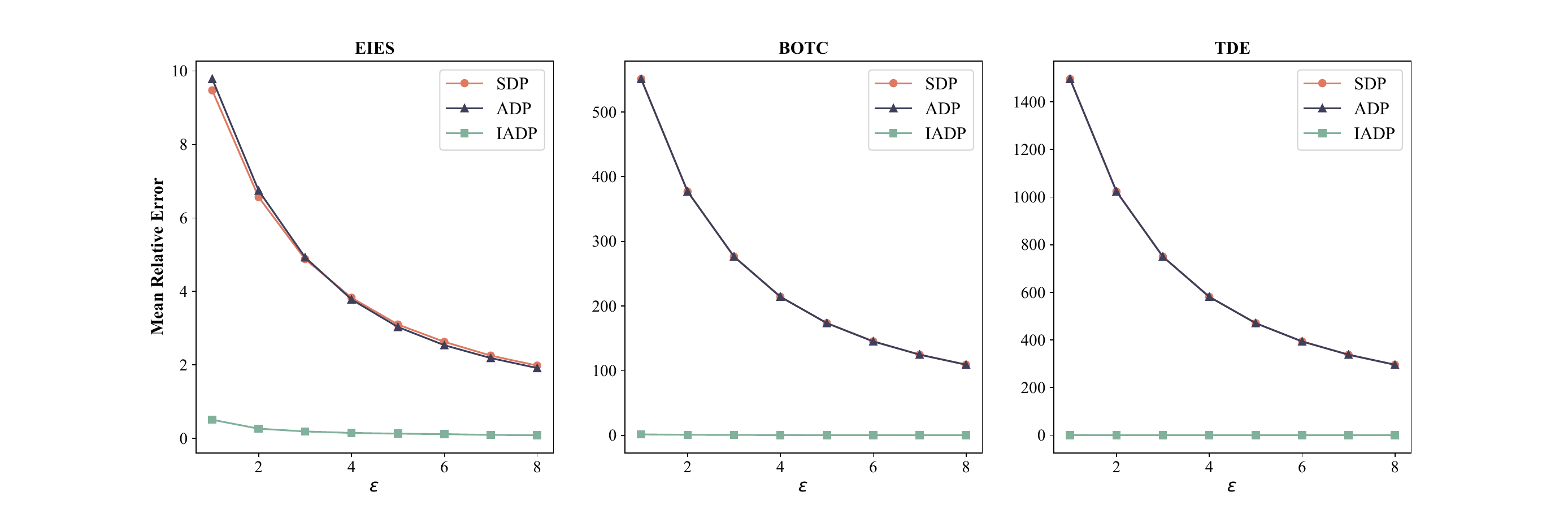}
    \caption{All-pair distance mean relative error for SDP, ADP and IADP under three real-world datasets EIES, BOTC and TDE with the $\varepsilon$ increases from $1$ to $8$.}
    \label{fig:add_result1}
\end{figure*}
\begin{figure*}
    \centering
    \includegraphics[width=1\linewidth]{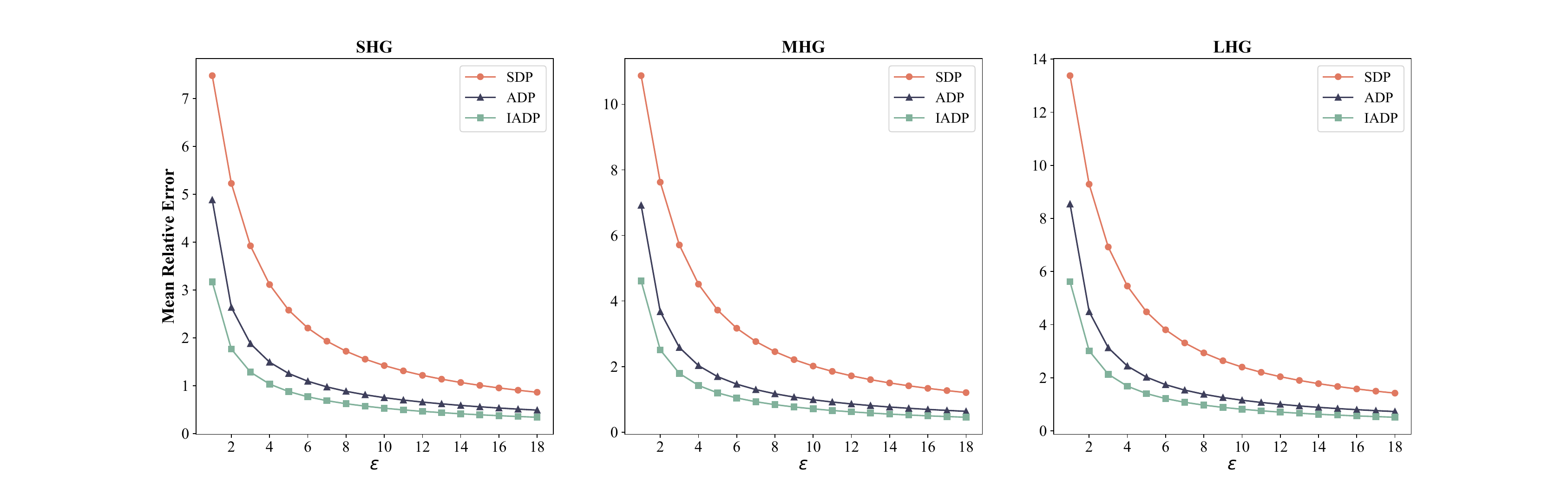}
    \caption{All-pair distance mean relative error for SDP, ADP and IADP under three synthesized datasets SHG, MHG and LHG with the $\varepsilon$ increases from $1$ to $18$.}
    \label{fig:remove_result1}
\end{figure*}
\begin{figure}
    \centering
    \includegraphics[width=0.8\linewidth]{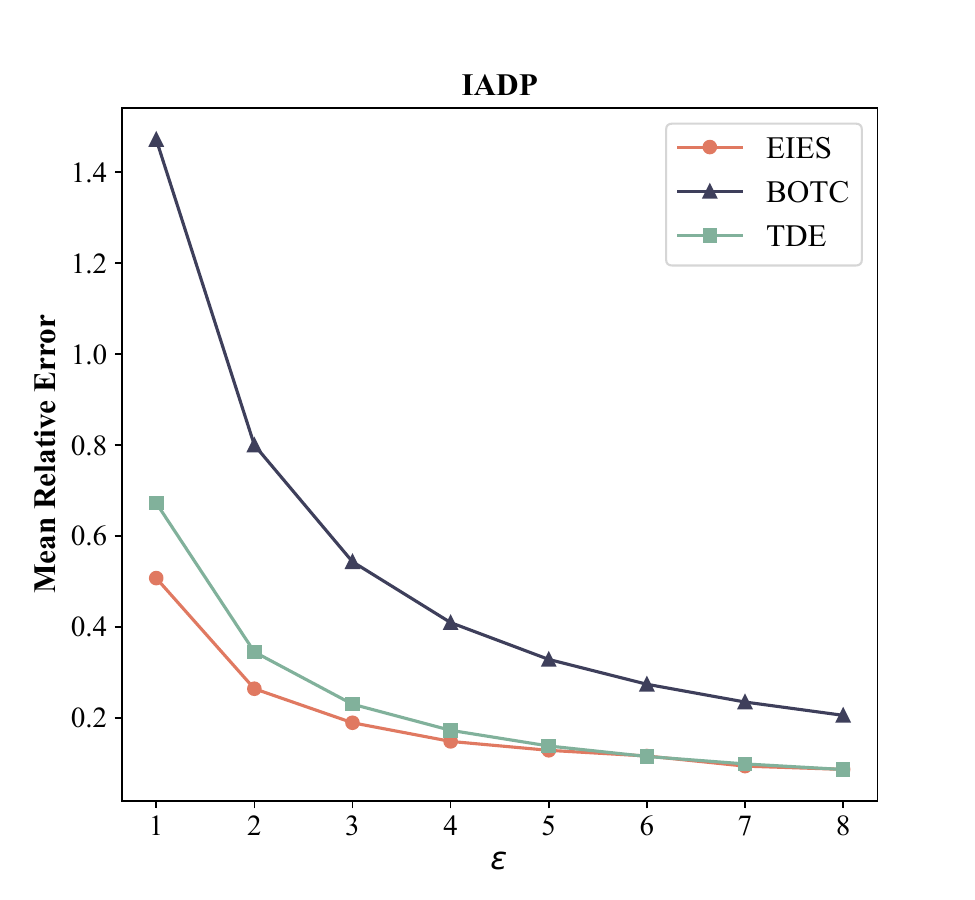}
    \caption{All-pair distance mean relative error for IADP under three real-world datasets EIES, BOTC and TDE with the $\varepsilon$ increases from $1$ to $8$.}
    \label{fig:add_result2}
\end{figure}
\begin{figure}
    \centering
    \includegraphics[width=0.8\linewidth]{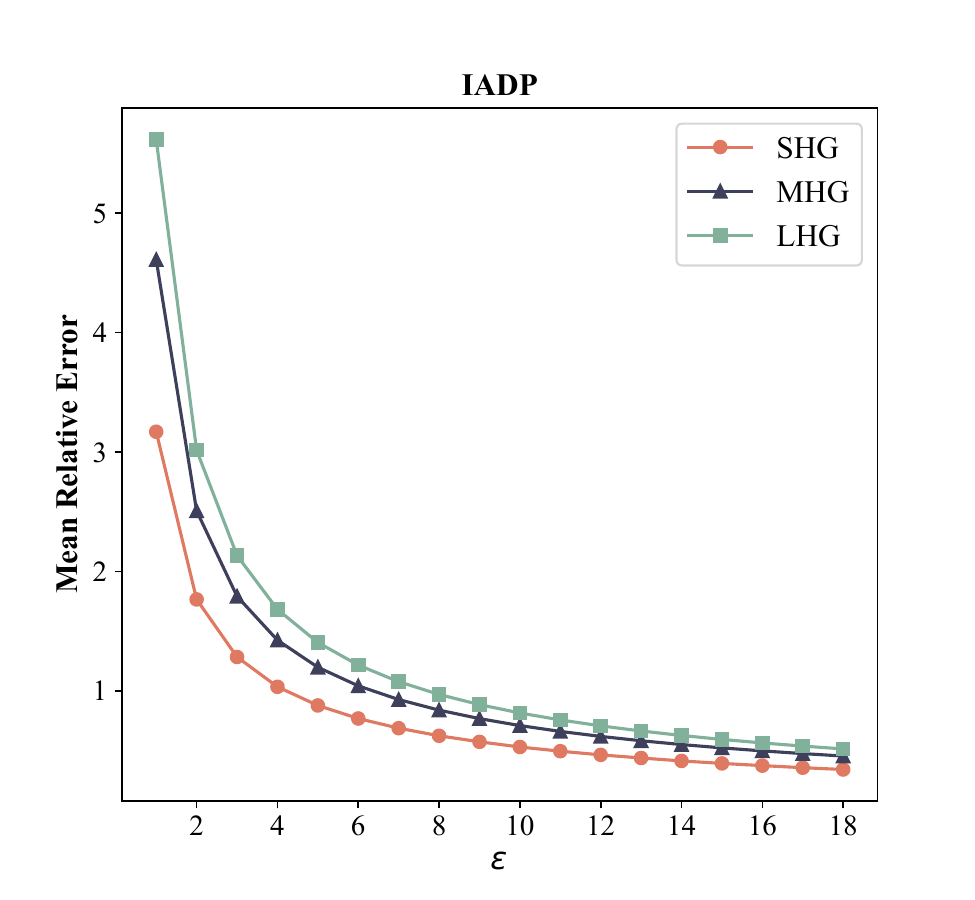}
    \caption{All-pair distance mean relative error for IADP under three real-world datasets SHG, MHG and LHG with the $\varepsilon$ increases from $1$ to $18$.}
    \label{fig:remove_result2}
\end{figure}
\subsection{Comparisons}
Fig. \ref{fig:add_result1} shows the all-pair distance MRE for SDP, ADP, and IADP under three real-world datasets EIES, BOTC and TDE with $\varepsilon$ increases from $1$ to $8$. For all datasets, IADP has a significant improvement in utility. Specifically, for the small graph EIES, IADP shows nearly a 10x improvement over SDP and ADP for small $\varepsilon$. For the larger graphs BOTC and TDE, IADP demonstrates over a 500x improvement compared to BOTC and TDE for small $\varepsilon$.
The significant increase in the utility of IADP over the other two solutions comes from the reduction in sensitivity, as demonstrated in the Table. \ref{tab:ds}. With $\varepsilon$ increases, this improvement decreases, which is due to the effect of $\varepsilon$ on the noise scale. The reason why SPD and ADP have close curves is that they share the same global sensitivity $n-1$. Due to the unbiased processing of ADP, the two exhibit similar MREs. It is worth noting that ADP has a smaller variance than SDP, and IADP shares that advantage \cite{kotsogiannis2020one}.

Fig. \ref{fig:add_result2} presents a more concrete comparison for IADP under three real-world datasets. The reason why BOTC has a larger error than either of the other two is that BOTC has a diameter of $9$, which is larger than both $2$ for EIES and $3$ for TDE. Thus, IADP can be effectively used in complex large-scale graphs. This is because our smooth sensitivity removes the dependence on $n$, and for denser graphs, the intuition is that the diameter will be smaller. The IADP improves the utility to an acceptable level. For $\varepsilon > 3$, MREs are smaller than $0.2$ for EIES and TDE. At $\varepsilon = 8$, the MREs for EIES and TDE are 0.0865 and 0.0862, respectively, indicating the high utility of our query results.

Fig. \ref{fig:remove_result1} demonstrates the all-pair distance MRE for SDP, ADP and IADP under three synthesized datasets SHG, MHG, and LHG with $\varepsilon$ increases from $1$ to $18$. With increasing $\varepsilon$, we consistently observe the MRE relation: SDP $>$ ADP $>$ IADP. There is a small difference from Fig. \ref{fig:add_result1} where the curves for SDP and ADP barely overlap. The smooth sensitivity for the datasets SHG, MHG, and LHG is $62$, $322$, and $1622$, respectively, leading to the addition of excessive negative noise to some query results. Truncation is employed to crop the negative noise values, reducing the overall error. Furthermore, due to the larger smooth sensitivity, we adjusted $\varepsilon$ to range from $1$ to $18$.
For $\varepsilon > 15$, we have that MRE is less than $1$.

Fig. \ref{fig:remove_result2} illustrates the MREs of IADP under SHG, MHG, and LHG.  For $\varepsilon > 4$, the MRE of the SHG is less than $1$. Specifically, at $\varepsilon = 9$, the MREs are $ 0.530$, $0.709$, and $0.815$ for SHG, MHG, and LHG, respectively, indicating reduced utility with a large sensitivity. Furthermore, at $\varepsilon = 18$, smaller MREs of $0.341$, $0.454$, and $0.514$ are achieved for SHG, MHG, and LHG, respectively.

\section{Discussion}\label{sec:8}
In this section, we present the limitation of ADP. Then, we discuss some potential issues that exist in our work, including utility of our results, strict assumption of graphs, and sensitivity of distance queries.
\subsection{Limitation of ADP}
$ALap_{\varepsilon,f}(x)$ is the implementation of ADP with exponential noise, however, it should satisfy $(\varepsilon,\delta)$-DP if query $f$ is monotonic over $p$-neighbors. Let us suppose that $x$ and $y$ are $p$-neighbors, that is, we have $f(x)\leq f(y)$. Given $z$, a random variable is drawn from the distribution in Definition \ref{def:Alap}. For $\forall S \subseteq \left[f(x),f(y)\right]$, we have $Pr\left[ ALap_{\varepsilon,f}(y)\subseteq S \right]=0$ while $Pr\left[ ALap_{\varepsilon,f}(x)\subseteq S\right]>0$. Thus, the indistinguishability of the outputs cannot be guaranteed by privacy loss $\varepsilon$. Furthermore, to maintain the DP guarantee, $ \min {\left(\delta\right)}$ should be $\varepsilon \backslash GS(f)$, which is unacceptable unless $\varepsilon$ is very small. Therefore, we reexamine the monotonicity of the query with respect to exponential noise (and its symmetric form), allowing the IADP to be free from its dependence on $\delta$.
\subsection{Limitation of IADP}
\subsubsection{Utility}

With the results in Section \ref{sec:7}, our solution has favorable utility on three real-world datasets. However, as depicted in Fig. \ref{fig:remove_result1} and Fig. \ref{fig:remove_result2}, although Alg. \ref{alg:2} significantly improves utility on Harary graphs, it still introduces an error of at least $34\%$ when $\varepsilon$ is $18$. Furthermore, reducing this error would require consuming more $\varepsilon$, which is intolerable in DP. In fact, we can see that the utility gap between our two experiment settings comes from the smooth sensitivity. For smooth sensitivity, if the neighbors are obtained by adding an edge, it equals the local sensitivity, which is very small. However, if the neighbors are obtained by removing an edge, the smooth sensitivity is also equal to the local sensitivity, which is very large. The reasons are different, the former is theoretically proven, while the latter is because $\max\{\Phi\}$ is the same as $\max\{\Psi\}$. Thus, denser or more connected Harary graphs have higher utility for distance queries.

\subsubsection{Strict Assumption}
For adding an edge, we assume that the graph $G$ is connected. For removing an edge, we need $G$ is $3$-edge connected, which is a strict assumption. In the real world, connected social graphs are very common, but graphs with connectivity of 3 are rare. Therefore, we only use synthetic graphs to validate our algorithm. Harary graphs are a class of graphs that are used to design reliable networks. Alg. \ref{alg:2} is suitable for distance queries oriented to reliable networks. To extend the applicability of this algorithm, we need to address the question of how to handle the case where the query result is $\infty$. If we answer $\infty$, privacy is exposed. However, refusing to answer can reveal privacy itself.

\subsubsection{Sensitivity}
One of the contributions of our work is to solve the problem of computing smooth sensitivities in distance queries. However, the smooth sensitivity does not differ by more than an order of magnitude from the global sensitivity on the synthetic graphs. When there exists a partial query whose smooth sensitivity has no significant advantage over the global sensitivity, it is appropriate to think about whether to apply smooth sensitivity. Despite the potential privacy issues with local sensitivity, we are still unable to fully prove the existence of the issue on the graphs. Therefore, it is still a topic worth studying whether local sensitivity can be applied on graph-related queries. This is a trade-off between privacy and utility.

\section{Conclusion and Future Work}\label{sec:9}
In this paper, we explore the asymmetric neighborhood setting in DP to answer distance queries with improved utility. We revisit the neighborhood definitions for iDP, OSDP, and ADP. On the basis of these works, we formally propose the definition of asymmetric neighborhood and asymmetric Laplace mechanism. Recognizing the potential privacy issue associated with local sensitivity, we integrate smooth sensitivity into our mechanism. Then, to privately publish the distance, we propose the solutions to calculate smooth sensitivity and publish the distance with a lower computational complexity in two neighbor operations: adding an edge and removing an edge. Finally, we use six datasets to validate our solutions.

The application of differential privacy to graph queries is significantly limited due to the complexity of graphs. Thus, our approach is to limit the sensitivity of specific queries to reduce noise damage to the utility. Our privacy preservation is achieved by masking the distance itself. An alternative strategy is to mask the edges, allowing us to adjust the distribution of edges on the graph while preserving the utility of distance. Another direction is to address the issue of refusal to answer caused by disconnections. The resolution of this issue could broaden the application areas of our solutions. Notably, we observe that for certain queries, their smooth sensitivity is equal to the local sensitivity. In this case, how to distinguish the privacy guarantees of local and smooth sensitivity remains an issue.

\bibliographystyle{IEEEtran}
\bibliography{references}{}

\end{document}